\documentclass[runningheads,a4paper]{llncs}

\usepackage{amsmath,amssymb}
\usepackage{graphicx}
\usepackage{newclude}
\usepackage{url}
\usepackage[utf8]{inputenc}
\usepackage[T1]{fontenc} 
\usepackage{caption} 
\usepackage{makecell}
\newcommand{\keywords}[1]{\par\addvspace\baselineskip
\noindent\keywordname\enspace\ignorespaces#1}

\begin{document}

\mainmatter  



\title{Content-Based Video Retrieval in Historical Collections of the German Broadcasting Archive}

\titlerunning{Content-based Video Retrieval in the German Broadcasting
Archive}

%
%
\author{Markus Mühling\inst{1} \and Manja Meister\inst{4} \and Nikolaus
Korfhage\inst{1} \and Jörg Wehling\inst{4} \and \\
Angelika Hörth\inst{4} \and Ralph Ewerth\inst{2,3} \and Bernd
Freisleben\inst{1}}

\authorrunning{Mühling et al.}

\urldef{\mailsa}\path|{muehling,korfhage,freisleb}@informatik.uni-marburg.de|
\urldef{\mailsb}\path|ralph.ewerth@tib.eu|
\urldef{\mailsc}\path|{manja.meister,joerg.wehling,angelika.hoerth}@dra.de|

\institute{Department of Mathematics and Computer Science, 
University of Marburg, \\
Hans-Meerwein-Str. 6, D-35032 Marburg, Germany \\
\mailsa\\ [2mm]
\and
German National Library of Science and Technology (TIB), \\
Welfengarten 1B, D-30167 Hannover, Germany\\
\mailsb\\ [1mm]
\and
Faculty of Electrical Engineering and Computer Science, \\
Leibniz Universität Hannover, \\ 
Appelstr. 4, D-30167 Hannover, Germany \\ [1mm]
\and
German Broadcasting Archive, \\
Marlene-Dietrich-Allee 20, D-14482 Potsdam, Germany\\
\mailsc\\
}

%
%

\maketitle

\begin{abstract}
The German Broadcasting Archive (DRA) maintains the cultural
heritage of radio and television broadcasts of the former German Democratic Republic (GDR).
The uniqueness and importance of the video material stimulates
a large scientific interest in the video content. 
In this paper, we present
an automatic video analysis and retrieval system for searching in
historical collections of GDR television recordings. It consists of video analysis algorithms for shot
boundary detection, concept classification, person recognition, text recognition
and similarity search. The performance of the system is evaluated 
from a technical and an archival perspective on 2,500 hours of GDR television
recordings.

\end{abstract}

\section{Introduction}
\label{sec:intro}

Digital video libraries become more and more important due to the new potentials
in accessing, searching and browsing the data
\cite{christel1995informedia,marchionini2002open,albertson2015design}.
In particular, content-based analysis and retrieval in large collections of
scientific videos is an interesting field of research. Examples are
Yovisto\footnote{http://www.yovisto.com},
ScienceCinema\footnote{http://www.osti.gov/sciencecinema}
and the TIB|AV-portal\footnote{http://av.tib.eu} of the German National
Library of Science and Technology (TIB).
The latter provides access to scientific videos based on speech recognition,
visual concept classification, and video OCR (optical character recognition).
The videos of this portal stem from
the fields of architecture, chemistry, computer science, mathematics, physics,
and technology/engineering.
 
The German Broadcasting Archive (DRA) in Potsdam-Ba\-bels\-berg provides access
to another valuable collection of scientifically relevant videos.
It encompasses significant parts of the audio-visual tradition in Germany
and reflects the development of German broadcasting before 1945 as well as
radio and television of the former German Democratic Republic (GDR).
The DRA was founded in 1952 as a charitable foundation and joint institution of the
Association of Public Broadcasting Corporations in the Federal Republic of
Germany (ARD).
In 1994, the former GDR's radio and broadcasting archive was established. The
archive consists of the film documents of former GDR television productions from
the first broadcast in 1952 until its cessation in 1991, including a
total of around 100,000 broadcasts, such as
contributions and recordings of the daily news program ``Aktuelle Kamera'',
political magazines such as ``Prisma'' or ``Der schwarze Kanal'',
broadcaster's own TV productions including numerous films, film adaptations and
TV series productions such as ``Polizeiruf 110'',
entertainment programs (``Ein Kessel Buntes''), children's and youth
programs (fairy tales, ``Elf 99'') as well as advice and sports programs.
Access to the archive is granted to scientific, educational and cultural
institutions, to public service broadcasting companies and, to a limited extent,
to commercial organizations and private persons. The video footage is often
used in film and multimedia productions.
Furthermore, there is a considerable international research
interest in GDR and German-German history. Due to the uniqueness and importance
of the video collection, the DRA is the starting point for many scientific studies.


To facilitate searching in videos, the DRA aims at digitizing and indexing the entire
video collection.
Due to the time-consuming task of manual video labeling, human
annotations focus on larger video sequences and contexts.
Furthermore, finding similar images in large multimedia archives is manually
infeasible. 

In this paper, an automatic video analysis and retrieval system for searching in
historical collections of 
GDR television recordings is presented.
It consists of novel algorithms for
visual concept classification, similarity search, person and text recognition 
to complement human annotations and to support users in
finding relevant video shots.
In contrast to manual annotations, content-based video analysis algorithms provide a more fine-grained analysis,
typically based on video shots.  
A GDR specific lexicon of 91 concepts including, for example,
``applause'', ``optical industry'', ``Trabant'', ``military parade'', ``GDR
emblem'' or ``community policeman'' is used for automatic annotation.
An extension of deep convolutional neural networks (CNN)
for multi-label concept classification, a comparison of a Bag-of-Visual Words
(BoVW) approach with CNNs in the field of
concept classification and a novel, fast similarity search approach are presented.
The results of automatically annotating 2,500 hours of GDR television recordings are
evaluated from a technical and an archival perspective.

The paper is organized as follows: Section
\ref{sec:cbvr} describes the video retrieval system and 
the video analysis algorithms. 
In Section \ref{sec:application}, experimental results 
on the GDR television recordings are presented. Section
\ref{sec:conclusions} concludes the paper and outlines areas for future research.


\section{Content-Based Video Retrieval}
\label{sec:cbvr}


\begin{figure}[tbp] 
	\centering
	\includegraphics[width=0.8\columnwidth]{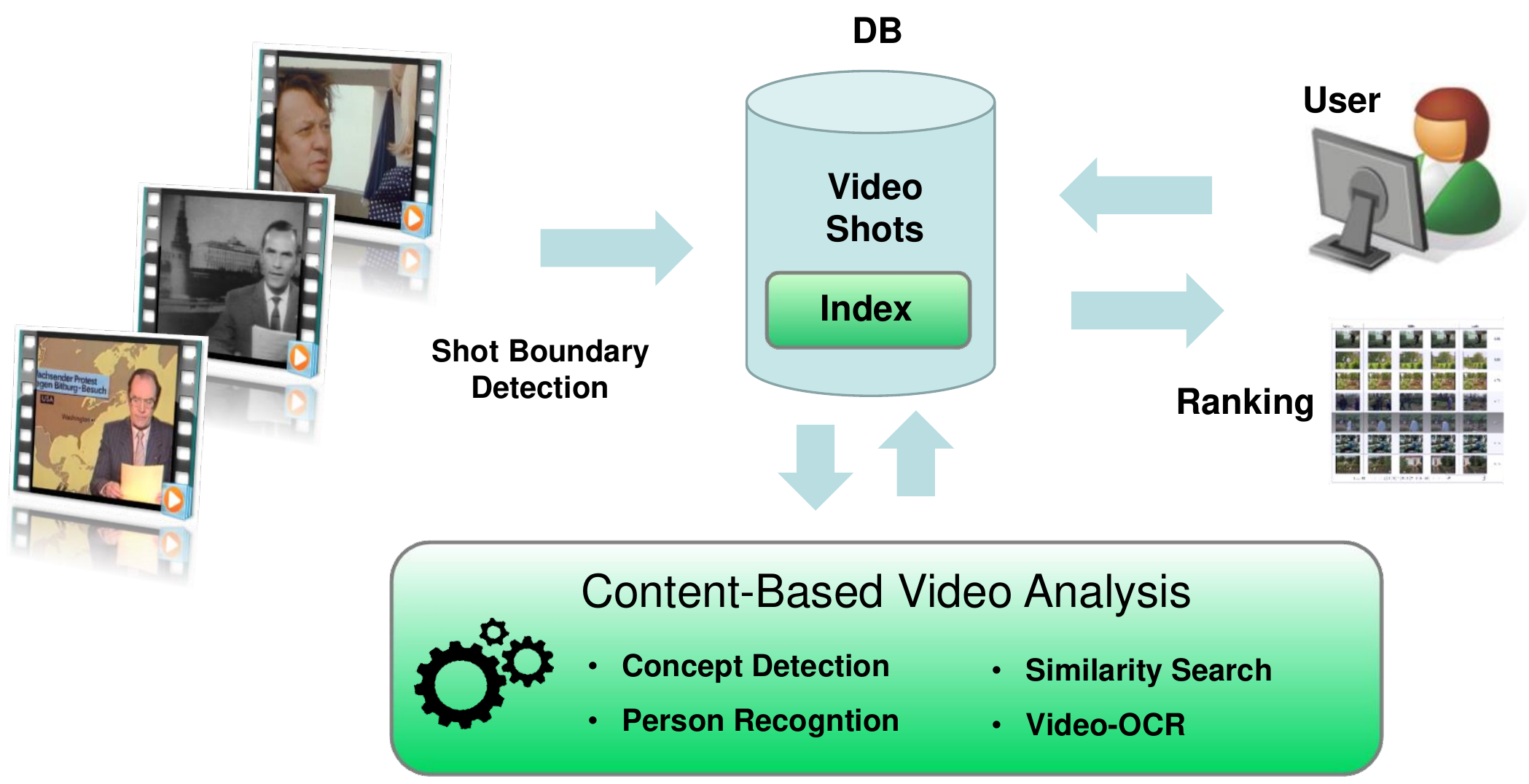}
	\caption[Video Retrieval System]{Video retrieval system.}
	\label{fig:retrieval_system}
\end{figure}

The aim of content-based video retrieval is to automatically assign semantic tags to video
shots for the purpose of facilitating content-based search and navigation.
Figure \ref{fig:retrieval_system} shows an overview of our developed video retrieval
system. First, the videos are preprocessed. This step mainly consists of shot
boundary detection \cite{ewerth2004video} in conjunction with thumb generation for the keyframes, which are required for later visualization purposes. 
The aim of shot boundary detection is the temporal segmentation
of a video sequence into its fundamental units, the shots. 
Based on video segmentation, the following automatic content-based video
analysis algorithms are applied: concept classification, similarity search, person
and text recognition.
The resulting metadata are written to a database and serve as an
intermediate description to bridge the ``semantic gap'' between the
data representation and the human interpretation.
Given the semantic index, arbitrary search queries can be processed
efficiently, and the query result is
returned as a list of video shots ranked according to the probability that the
desired content is present. In the following, the content based video analysis algorithms are described in
more detail. 


\subsection{Visual Concept Classification}
\label{subsec:concepts}

The classification of visual concepts is a challenging task due to the large
complexity and variability of their appearance. Visual concepts can be, for example,
objects, sites, scenes, personalities, events, or activities. 
The definition of the GDR specific concept lexicon is based on the analysis of
user search queries with a focus on queries that were experienced as difficult and time-consuming to answer manually. 
A lexicon of 91 concepts was defined after analyzing more than 36,000 user queries received within a five-year period from 2008 to 2013. Recent user queries, assumed to be of future research interest, were summarized thematically and ordered by frequency. The concept lexicon comprises events such as ``border control'' and ``concert'', scenes such as
``railroad station'' and ``optical industry'', objects like ``Trabant'' or 
activities such as ``applauding''.  
Manually annotated training data was used to build the concept models. For
this purpose, a client-server based annotation tool was built to facilitate
the process of training data acquisition and to select a sufficient quantity of
representative training examples for each concept. We defined a minimum number of 100 positive samples per concept. 


Recently, deep learning algorithms fostered a renaissance of artificial neural networks,
enabled by the massive parallel processing power of modern graphics cards. 
Deep learning approaches, especially deep CNNs, facilitated
breakthroughs in many computer vision fields
\cite{Krizhevsky2012,Graves2013,Breuel2013,deepface}.
Instead of using hand-crafted features such as SIFT descriptors
\cite{Lowe1999}, CNNs learn the features automatically during the training process. 
A CNN consists of several alternating convolution and max-pooling layers with
increasingly complex feature representations and typically has 
several fully connected final layers.

State-of-the-art network architectures for image recognition
\cite{Krizhevsky2012, Szegedy2014} as well as the current 
datasets \cite{Deng2009, Zhou2014} consider only a single concept per image
(``single-label''). In contrast, real world concept classification scenarios are
multi-label problems. Several concepts, such as  ``summer'', ``playground'' and ``teenager'', may
occur simultaneously in an image or scene. While some approaches use
special ranking loss layers \cite{gong2013deep}, we have extended the CNN
architecture using a sigmoid layer instead of the softmax layer and a
cross entropy loss function.

Furthermore, millions of training images are needed to build a deep CNN model
from scratch. Due to the relatively small amount of available training data, we
have adapted a pre-trained CNN classification model (GoogleNet \cite{Szegedy2014}
trained on ImageNet \cite{Deng2009}) to the new GDR concept lexicon using our
multi-label CNN extension and performed a fine-tuning on the GDR television recordings.
The models were trained and fine-tuned using the deep learning
framework Caffe \cite{Jia2014}. 
In addition, a comparison between a state-of-the-art
BoVW approach and our deep multi-label CNN was performed on the publicly available, fully
annotated NUSWIDE scene dataset \cite{Chua2009}. 
The BoVW approach is based on a combination of different
feature representations relying on optimized SIFT variants \cite{muehling2014thesis}. The different
feature representations are combined in a support vector machine (SVM)-based
classifier using multiple kernel learning \cite{muehling2012multimodal}.
Our deep multi-label CNN approach significantly outperformed the
BoVW approach by a relative performance improvement
of almost 20\%. 
In contrast to binary SVM classifiers, deep neural networks are
inherently multi-class capable so that only a single compact model has to be built for
all concept classes.
While the runtime of the BoVW approach takes already 1.97 seconds for
feature extraction and the classification runtime depends linearly on the number
of concepts, our deep multi-label CNN takes only 0.35 seconds on the CPU (Intel Core
i5) and  just 0.078 seconds on the GPU (Nvidia Titan X).
Although deep neural networks are computationally expensive in the training phase, 
they are very efficient at the classification stage.
Altogether, deep multi-label CNNs provide better recognition quality, compact
models and faster classification runtimes.

\subsection{Similarity Search}
\label{subsec:similaritysearch} 

Since the DRA offers researchers a large number of video recordings containing
several millions of video shots, the need for a system that helps to rapidly 
find desired video shots emerges. While scanning through the whole
video archive is practically infeasible for humans, a possible solution is
to index the videos via concepts as described in Section
\ref{subsec:concepts}.
However, this approach requires manually annotated training images for learning
the concept models. Additionally, search queries are restricted to the
vocabulary of predefined concepts, and new concept models have to be developed on
demand.
In contrast to textual concept-based queries, image-based queries provide users
more flexibility and new ways of searching.

While query-by-content based on low-level features turned out to be insufficient
to search successfully in large-scale multimedia databases,
image representations learned by deep neural networks greatly improved the
performance of content-based image retrieval systems \cite{Wan2014}. 
They are less dependent on pixel intensities and are clearly better suited for
searching semantic content. 
However, high-dimensional CNN features are not well suited for searching efficiently in large video
collections. Fast search in large databases is an essential requirement for
practical use. For this purpose, proposals for learning binary image codes for compact
representations and fast matching of images have been made. Krizhevsky and Hinton \cite{Krizhevsky2011}, for
example, used very deep autoencoders, and Lin et al. \cite{Lin2015} extended a CNN to
learn binary hash codes for fast image retrieval. 

\begin{figure}[tbp] 
	\centering
	\includegraphics[width=0.9\columnwidth]{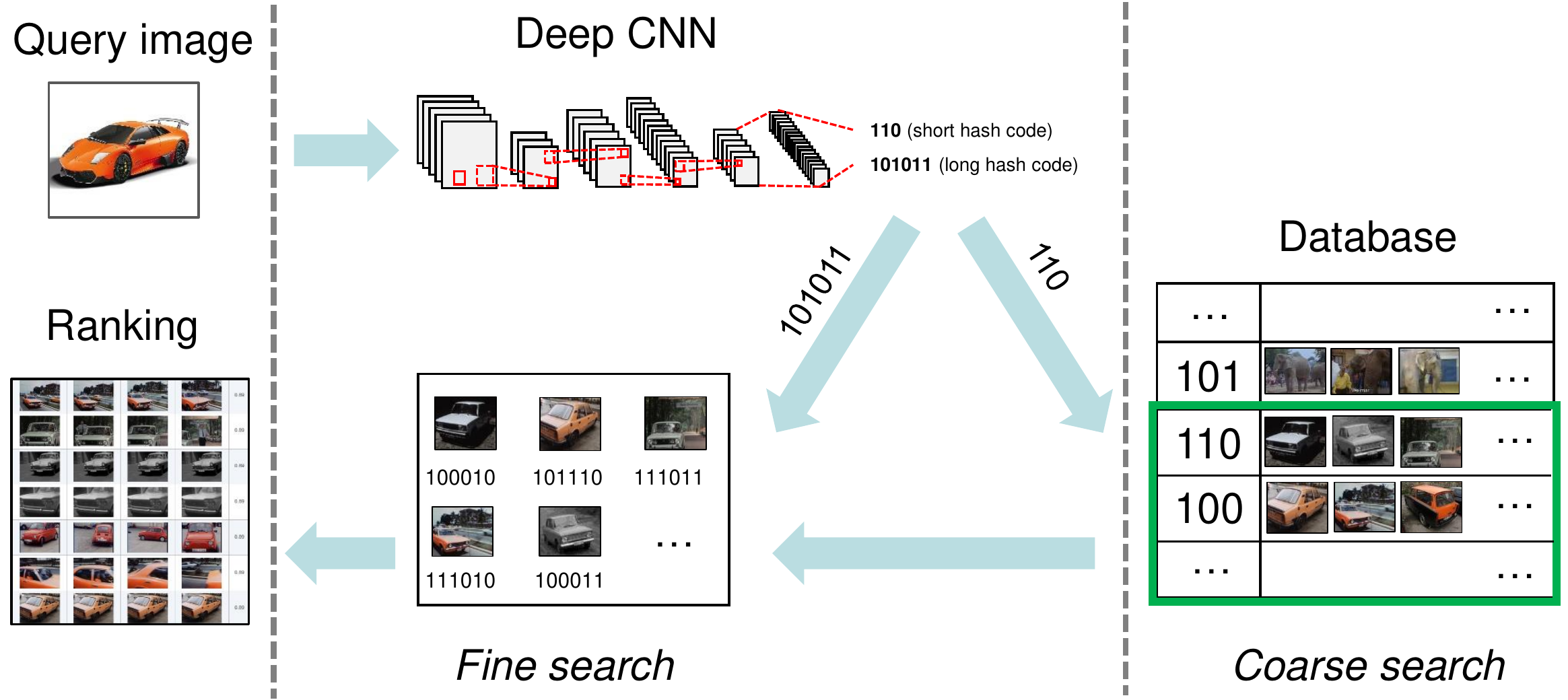}
	\caption[Content based similarity search]{Content based similarity search.}
	\label{fig:similarity_search}
\end{figure}

In this section, an approach for fast content-based similarity search in
large video databases is presented. 
For efficient storage and matching of images, binary codes are learned by
deep CNNs. 
This mapping of images to binary codes is often referred to as ``semantic
hashing'' \cite{Salakhutdinov2009}.
The idea is to learn a ``semantic hash function'' that maps similar images to
similar binary codes.
To learn the hash function, a method similar to the approach described by Lin et al. \cite{Lin2015} is used. Based on a pre-trained CNN classification
model, we devised a coding layer and an appropriate loss layer for error
propagation for the network architecture. The best results were obtained using Chatfield et al.'s network \cite{Chatfield2014} trained on the PLACES dataset \cite{Zhou2014}.
An advantage of using pre-trained classification models is the speed-up
in training time.
To obtain high-level features, we built the coding
layer on top of the last fully-connected layer. Furthermore, the hash function
can be adapted to unlabeled datasets by using the predictions of the 
pre-trained classification model for error propagation.

The overall video retrieval system is based on the analysis of 
keyframes, i.e., representative images.  In our approach, five frames (the first, the last, and three in between) per video shot are used as keyframes for indexing.
Given the hash function, the keyframes of the video collection are
fed into the deep CNN and the mapped binary codes are stored in the
database. Based on the resulting index, queries-by-image can be answered by
matching the binary code of the given image to the database. The
overall retrieval process is shown in Figure \ref{fig:similarity_search}. Given
the query image, the binary codes are extracted using the learned deep CNN. We
use a two-stage approach based on semantic hashing. First, a coarse search is
performed using 64-bit binary codes resulting in a ``short'' list of potential
results. The hamming distance is applied to compare the binary codes, and a
vantage point tree \cite{Yianilos1993} is used as an additional index structure to accelerate the search process. The resulting short list consists of the 10,000 nearest neighbors.
The longer the binary codes, the more accurate the image representations are.
Therefore, in the second stage, a refined search using 256-bit binary codes is
performed on the short list, and the images are ranked according to the distance to the query image. Differently from Lin et al. \cite{Lin2015}, hash codes are used for the refined search as well. In our two-stage approach, both coding layers,
for 64-bit and 256-bit binary codes, are integrated into the same architecture
resulting in a concurrent computation at training and testing time.
Finally, the resulting images are mapped to video shots. 



\subsection{Person Recognition}
\label{subsec:persons}


Based on the analysis of user search queries, the GDR specific concept lexicon
has been extended by 9 personalities, including ``Erich Honecker'', ``Walter
Ulbricht'', ``Hilde Benjamin'', ``Siegmund Jähn'', ``Hermann Henselmann'',
``Christa Wolf'', ``Werner Tübke'', ``Stephan Hermlin'' and ``Fritz Cremer''.
Instead of using concept classification, personalities are handled using a face
recognition approach. Therefore, feature representations of known persons are
stored in a face database, and a face recognition system was
built that scans the video shots and recognizes the identity of a detected face
image by comparing it to the face database.
Finally, the resulting index of person occurrences can be used in search
queries.
The face processing pipeline consists of several
components: face detection, face alignment and face recognition.
For the face recognition component, we evaluated the following approaches:
Fisherfaces \cite{belhumeur1997fisherfaces}, Local Binary Pattern Histograms
\cite{ahonen2004lbp} and a commercial library, called FaceVACs\footnote{http://www.cognitec.com}. 
Furthermore, we evaluated whether
a preprocessing step using grayscale histogram equalization, training data
augmentation using Google search queries, or a face
tracking component improve the recognition accuracy. Based on the results of our evaluations, we finally
used the method of Viola and Jones \cite{viola2001} for face
detection and FaceVACs for face alignment and
recognition.

\begin{figure}[tb] 
	\centering
	\includegraphics[width=0.99\columnwidth]{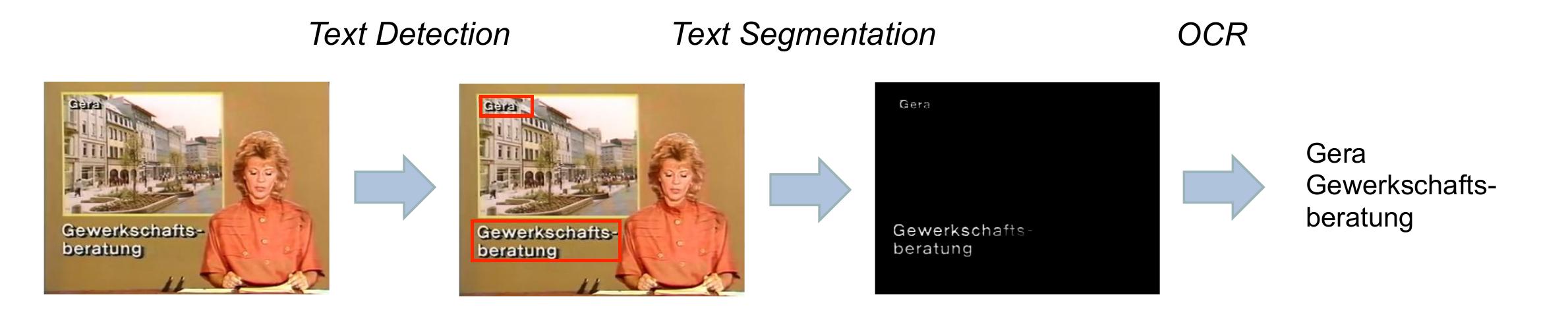}
	\caption[OCR]{Text recognition pipeline.}
	\label{fig:ocr}
\end{figure}

\subsection{Text Recognition (Video OCR)}
\label{subsec:ocr}

Superimposed text often hints at the content of a video shot. In news videos,
for example, the text is closely related to the current report, and in silent
movies it is used to complement the screen action with cross headings. Involved
algorithms can be distinguished by their objective, whether it is text
detection, also called localization, text segmentation, or optical character recognition (see Figure \ref{fig:ocr}).

We developed a text recognition system that allows users to search for in-scene
and overlaid text within the video archive. 
For this purpose, the I-frames of the videos are analyzed and the recognized ASCII text
is stored in the database. Based on the resulting index, OCR search
queries can be answered by a list of video shots ranked according to the
similarity to the query term.
Due to poor technical video quality and low contrast of text appearances, the similarities 
between the query term and the words in the database are calculated
using the Levenshtein distance.

For text detection, localization and segmentation in video frames, a method
based on Maximally Stable Extremal Regions (MSER) \cite{matas2004robust} is used. It is able to
detect both overlaid text, as well as text within the scene, for example on banners.
Experimental results revealed that the text segmentation component plays
an important role in the case of poor quality videos. 
Text segmentation crops localized text out of the image to yield
black letters on a white background. This step is necessary to feed the result
into an OCR algorithm that transforms the image into machine-readable text. A
non-uniform background would normally impair this process.
For OCR, we finally used the open source library 
Tesseract\footnote{http://code.google.com/p/tesseract-ocr/}.

\section{Experimental Results}
\label{sec:application}

\begin{figure}[tbp] 
	\centering
	\includegraphics[width=0.8\columnwidth]{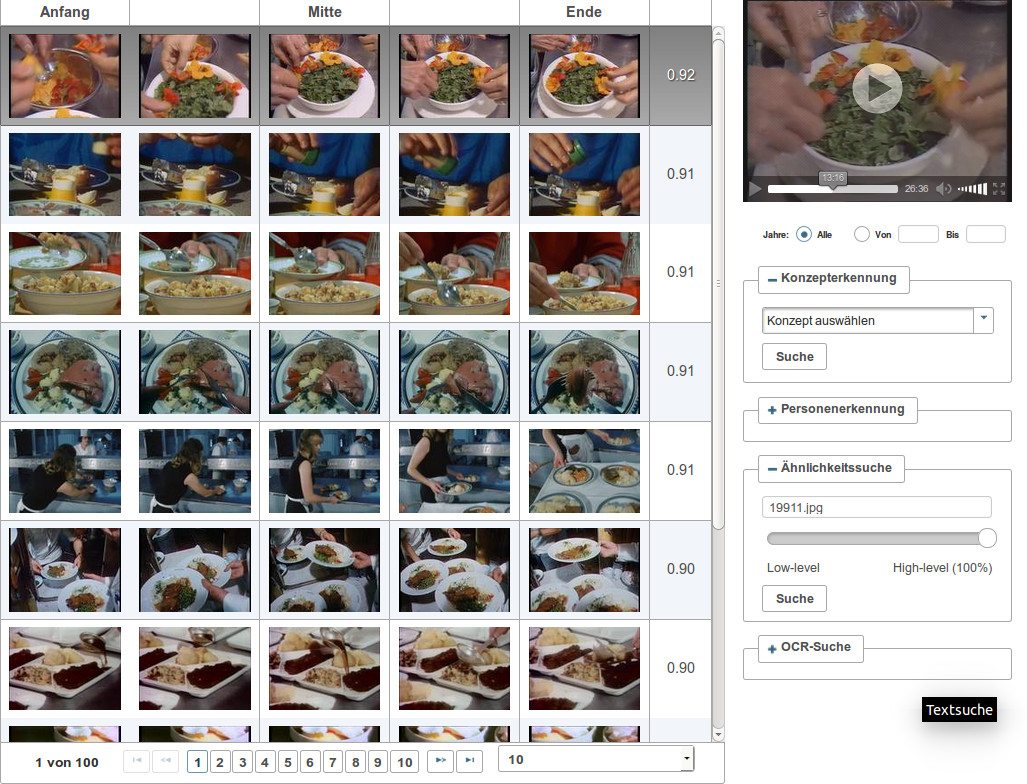}
	\caption[Ranking]{Retrieval results for a query image showing a meal.}
	\label{fig:ranking}
\end{figure}

A web-based GUI has been developed to automatically respond to user
queries related to concepts, persons, similar images and
text. The retrieval results are presented to the user in the form of a ranked list
of video shots (see Figure \ref{fig:ranking}) where each video shot is
represented by five video frames and a probability score indicating the
relevance.
Furthermore, a video player allows to visually inspect the video shots.

\subsection{Data}
In total, more than 3,000 hours of historical GDR television recordings have been
digitized by now. The video footage is technically very challenging. Many recordings are in
grayscale and of low technical quality; the older the recordings, the poorer the video
quality. The temporal segmentation of the videos resulted in approximately 2
million video shots. From these shots, 416,249 have been used
for the training process and 1,545,600 video shots, corresponding to about 2,500
hours of video data, for testing.



\subsection{Results}

In the following, the results for concept classification, person
recognition, similarity search and video OCR are presented.
The results are evaluated using the average precision (AP) score that is the
most commonly used quality measure in video retrieval. The AP score is
calculated from the list of video shots as follows:
\begin{equation}
AP(\rho)=\frac{1}{\left|R \cap \rho^N\right|}\sum_{k=1}^N \frac{\left|R \cap \rho^k\right|}{k} \psi(i_k)
\label{eq:fund:ap}\;\;\;\;
\textrm{with} \quad \psi(i_k) =
\begin{cases} 
\ 1 & \mbox{if } i_k \in R 
\\[5pt] \ 0  & \mbox{ otherwise } 
\end{cases}
\vspace{3pt}
\end{equation}
where $N$ is the length of the ranked shot list,
$\rho^k=\{i_1,i_2,\ldots,i_k\}$ is the ranked shot list up to rank $k$, $R$
is the set of relevant documents, $\left|R \cap \rho^k\right|$ is the number of
relevant video shots in the top-$k$ of $\rho$ and $\psi(i_k)$ is the relevance
function.
Generally speaking, AP is the average of the precisions at each relevant
video shot.
To evaluate the overall performance, the mean
AP score is calculated by taking the mean value of the AP scores from different
queries.

\subsubsection{Concept Classification and Person Recognition.}
In total, 86 concepts, comprising 77 concepts and 9 persons, were evaluated. 
Of the originally 91 concepts 14 were dismissed due to an insufficient number of training images. However, another 14 of the 77 evaluated concepts have less than 100 training images.   
Altogether, 118,020 positive training examples were gathered for learning
the concept models. The retrieval results for concepts and persons were
evaluated based on the top-100 and top-200 ranked video shots.
Although 14 concepts have less than 100 training images and despite poor video
quality, we obtained a mean AP of 62.4\% and 58.0\%, respectively. Even
concepts occurring predominantly in grayscale shots of low video quality
yielded good results, such as ``daylight mining'' with 84.3\% AP. These
results reveal the high robustness of the proposed multi-label deep CNN approach with respect to low
quality historical video data.

For person recognition, we achieved a very good result of 83.3\% mean AP on
the top-100 and 81.1\% mean AP on the top-200 video shots and even 100\% AP for
distinctive and frequently occurring personalities, such as Erich Honecker and
Walter Ulbricht. In summary, we achieved a mean AP of 64.6\% and 60.5\% on the top-100 and
top-200, respectively, for both concepts and persons.

\subsubsection{Similarity Search.}

The interpretation whether two images are similar is subjective and
context specific. 
The definition of similarity ranges from pixel-based similarity
to image similarity based on the semantic content.
How much low-level and semantic similarity contribute to the retrieval results
can be individually adjusted in the GUI. Furthermore, two use cases have been
implemented:
searching by video frames selected from within the corpus and searching by
external images, e.g., downloaded from the Internet.
In our evaluation, we focus on the more difficult task of semantic similarity
using 50 external query images from the Internet chosen collaboratively
by computer scientists and archivists.  An example result for a query
image showing a meal is presented in Figure \ref{fig:ranking}. Each retrieval
result was evaluated up to the first 100 video shots obtaining a mean AP of 57.5\%.
The image similarity search relies on representative images from the video
shots. Based on an image corpus of more than 7 million images, we achieved a very fast
response time of less than 2 seconds. 

 
 \subsubsection{Video OCR.}

For the task of text retrieval, 46 query terms according to
previously observed search query preferences of DRA users have been evaluated.
Based on these 46 search queries, like ``Abschaffung der Todesstrafe'',
``Kinder- und Jugendspartakiade'', ``Öffnungszeiten'', ``Protestbewegung'',
``Rauchen verboten'', ``Warschauer Vertrag'', ``Planerfüllung'', ``Nationale
Front'', ``Gleichberechtigung'', ``Mikroelektronik'' or ``Staatshaushalt'' a very satisfying
retrieval performance of 92.9\% mean AP has been obtained.
As expected, the results for overlaid text are
significantly better than for text within the scene.


\subsection{Archivist's Perspective} 

In this section, the presented content-based video retrieval system is
evaluated from an archivist's perspective with a focus on usability and
usefulness for archivists and DRA users.
Since users of the archive are often looking for everyday scenes in the former
GDR, concepts such as ``pedestrian'', ``supermarket'', ``kitchen'', ``camping
site'', ``allotment'' or ``production hall'',  in conjunction with the high search
quality for most of these concepts,  are a valuable contribution to
help researchers finding appropriate scenes. 
Concepts with an AP score of more than approximately 50\% turned out to be
practically very useful. 66\% of the concepts achieved an AP score of more than
50\%.

Searching manually for persons in videos is a quite time
consuming task, particularly for less known members of the Politbüro and 
Ministers of the GDR. Thus, the high quality of the provided automatic person
indexing algorithms is a great benefit for archivists as well as for users of the archive. 


The implemented similarity search system tremendously extends the accessibility
to the data in a  flexible and precise way. It provides complementary
search queries that are often hard to verbalize. In addition, it
allows an incremental search. Previous results may serve as a source of
inspiration for new similarity search queries for refining search intentions.

Another useful search option is offered by video OCR. OCR search results are very helpful since
overlaid text is often closely related to the video content. The
system recognizes the majority of slogans, locations and other terms precisely. 


Altogether, the fine-grained automatic annotation is a very valuable supplement
to human-generated meta-data.
Due to the variability of the content-based video retrieval system, different
user needs are taken into account. The combination of different search modalities allows to
answer a wide range of user queries leading
to more precise results in significantly less time.

\section{Conclusion}
\label{sec:conclusions}


In this paper, we have presented a content-based video retrieval system for 
searching in historical collections of GDR television recordings.
Novel algorithms for
visual concept classification, similarity search, person recognition and video OCR 
have been developed to complement human annotations and to support users in
finding relevant video shots.
Experimental results on about 2,500 hours of GDR television recordings have indicated the excellent
video retrieval quality in terms of AP as well as in terms of usability and
usefulness from an archivist's perspective.

There are several areas for future work. First, the concept-based approach
requires manually annotated training images for learning the concept models. 
New concepts have to be developed on demand. Therefore, it is
important to reduce the effort for training data acquisition. Second, the
audio modality can be used to improve the detection performance of several
audio related concepts. Finally, a similarity search query can be very
subjective depending on a specific user in given situation. New strategies have
to be developed to predict the user's intention.

\section{Acknowledgements}
This work is financially supported by the German Research Foundation
(DFG-Programm ``Förderung herausragender Forschungsbibliotheken'', ``Bild- und
Sze\-nenrecherche in historischen Beständen des DDR-Fernsehens im Deutschen
Rundfunkarchiv durch automatische inhaltsbasierte Videoanalyse''; CR 456/1-1, EW 134/1-1, FR 791/12-1).

\bibliographystyle{splncs03}
\bibliography{refs}

\end{document}